\documentstyle[12pt]{article}
\begin{document}
\begin{flushright}
DPNU-96-29\\
SNUTP-96-037
\end{flushright}
\vspace{5pt}
\setlength{\baselineskip}{0.25in}
\renewcommand{\thefootnote}{\fnsymbol{footnote}}
\begin{center}
{\Large{\bf Dynamical Symmetry Breaking in a Gauge Theory ``Thirring Model''
\footnote{To appear in the proceedings of the 14th Symposium on Theoretical 
Physics, Cheju, Korea,\\
\hspace*{15pt}
from 21 July to 26 July, 1995. Talk was given by Y. Kim.}
}}
\vspace{30pt}
\noindent

Taichi Itoh, Yoonbai Kim${}^{\dagger}$, Masaki Sugiura, Koichi Yamawaki
\vspace{10pt}
\noindent

Department of Physics, Nagoya University, Nagoya 464-01, Japan\\
taichi, sugiura, yamawaki@eken.phys.nagoya-u.ac.jp\\
$\mathop{}^{\dagger}$Department of Physics, Pusan National University,
Pusan 609-735, Korea\\
yoonbai@top.phys.pusan.ac.kr
\end{center}
\vspace{5mm}

\begin{abstract}
Dynamical fermion mass generation is studied in the three-dimensional 
Thirring model reformulated as a gauge theory by introducing hidden
local symmetry.
The analysis by use of Schwinger-Dyson equation is shown
to exhibit a critical behavior as the number $N$ of four-component fermions
approaches $N_{\rm cr}=128/3\pi^{2}$.
\end{abstract}

\vspace{5mm}

\def\fsl#1{\setbox0=\hbox{$#1$}           
   \dimen0=\wd0                                 
   \setbox1=\hbox{/} \dimen1=\wd1               
   \ifdim\dimen0>\dimen1                        
      \rlap{\hbox to \dimen0{\hfil/\hfil}}      
      #1                                        
   \else                                        
      \rlap{\hbox to \dimen1{\hfil$#1$\hfil}}   
      /                                         
   \fi}
\newcommand{\dfrac}[2]{\frac{\strut \displaystyle{#1}}%
{\strut \displaystyle{#2}}}

When we consider the models of massless self-interacting fermions, a central 
issue
is the dynamical fermion mass generation. If it is indeed the case,
intriguing detailed questions are whether the model of interest is shown to
exhibit a critical behavior as the number $N$ of fermions
approaches $N_{\rm cr}$ and, in three dimensions, whether the pattern of 
dynamically generated masses preserves the parity or not.

In this report we address ourselves to the problem of dynamical symmetry
breaking in the Thirring model of $N$ massless four-component fermions.
(For the notations and the detailed calculations, see Ref.\cite{IKSY}.)
\begin{equation}\label{thi}
{\cal L}_{\rm Thi}=\sum_{a}\bar{\psi}_{a}i\gamma^{\mu}\partial_{\mu}\psi_{a}
-\frac{G}{2N}\sum_{a,b}\bar{\psi}_{a}\gamma^{\mu}\psi_{a}\,\bar{\psi}_{b}
\gamma_{\mu}\psi_{b},
\end{equation}
where $\psi_{a}$ is a four-component Dirac fermion and the indices $a$, $b$
run over $N$ fermion species. Since the form of four-fermion interaction
term is a contact term between vector currents, a well-known technique
to facilitate $1/N$-expansion is to rewrite Eq.(\ref{thi}) by introducing 
an auxiliary vector field $A^{\mu}$ such as \cite{Par}$\sim$\cite{Han}
\begin{equation}\label{aux}
{\cal L}_{\rm aux}= \sum_{a}\bar{\psi}_{a}i\gamma^{\mu}(\partial_{\mu}
-\frac{i}{\sqrt{N}}A_{\mu})\psi_{a}+\frac{1}{2G}A_{\mu}A^{\mu}.
\end{equation}
Gauge-noninvariant as the above Lagrangian is, however one may be tempted
to regard $A_{\mu}$ as a gauge field. A systematic way to construct
the $U(1)$ gauge theory, but is gauge equivalent to Eq.(\ref{aux}) is 
to elicit the fictitious Goldstone degree (or equivalently the St\"{u}ckelberg
field) based on the principle of hidden local symmetry \cite{BKY},
\cite{KK,IKSY,Kon}:
\begin{equation}\label{hls}
{\cal L}_{\rm HLS}=\sum_{a}\bar{\psi}_{a}i\gamma^{\mu}D_{\mu}\psi_{a}+
\frac{1}{2G}(A_{\mu}-\sqrt{N}\partial_{\mu}\phi)^{2}.
\end{equation}
It is obvious that Eq.(\ref{hls}) possesses a $U(1)$ gauge symmetry and 
the gauge-fixed (unitary gauge) form of it exactly coincides with 
Eq.(\ref{aux}), so does the original Thirring model in Eq.(\ref{thi}). 

Now that we find a gauge-invariant formulation of the Thirring model, we
have the privilege to choose the gauge appropriate for our particular
purpose. Here, instead of the unitary gauge notorious for loop
calculations, let us consider the nonlocal $R_{\xi}$ gauge at the Lagrangian
level
\begin{equation}
\label{GF}
{\cal L}_{\rm GF}=-\frac{1}{2}\Bigl(\partial_{\mu}A^{\mu}+\sqrt{N} 
\frac{\xi(\partial^2)}{G}\phi\Bigr){1 \over \xi(\partial^2)}\Bigl(
\partial_{\nu}A^{\nu}+\sqrt{N}\frac{\xi(\partial^2)}{G}\phi\Bigr),
\end{equation}
where the gauge fixing parameter $\xi$ has the momentum- (derivative-)
dependence. It is straightforward to prove the possession of the BRS symmetry
in spite of the nonlocality of $\xi$ and thereby it guarantees the S-matrix
unitarity. Another intriguing point is that, in the combined Lagrangian of
Eqs. (\ref{hls}) and (\ref{GF}), the fictitious Nambu-Goldstone
boson $\phi$ is completely decoupled independently of the specific form of
$\xi(\partial^{2})$. 

In (2+1) dimensions, our Lagrangian in Eq.(\ref{hls}) is invariant under
the parity
\begin{equation}
\psi_{a}(x)\mapsto\psi_{a}^{'}(x^{'})=i\gamma^{3}\gamma^{1}\psi_{a}(x),
\;\;\;A_{\mu}(x)\mapsto A_{\mu}^{'}(x^{'})=(-1)^{\delta_{\mu 1}}A_{\mu}(x).
\end{equation}
and the so-called global ``chiral'' transformation
\begin{equation}
\psi_{a}\mapsto\psi^{'}_{a}=\Bigl(\exp\bigl(i\omega^{i\alpha}
\frac{\Sigma^{i}}{2}\otimes T^{\alpha}\bigr)\psi\Bigr)_{a},
\end{equation}
where $\Sigma^{0}=I,\;\Sigma^{1}=-i\gamma^{3},\;\Sigma^{2}=\gamma^{5},\;
\Sigma^{3}=-\gamma^{5}\gamma^{3}$ and $T^{\alpha}$ denote the generators 
of $U(N)$.
The question we shall address from now on is ``which symmetry is broken 
dynamically?'' In concern with the parity, first issue is whether one can
take the regularization to keep both the $U(1)$ gauge symmetry and 
the parity or not. Since our gauge action in tree level has
the parity-conserving mass and the number of two component Dirac fermion 
species is even, the parity need not be violated by appropriate regulator.
For example, the introduction of parity-conserving Pauli-Villars regulator 
leads to the parity-invariant effective action for the gauge field as have done
in (2+1)D quantum electrodynamics(QED${}_{3}$) \cite{Red}. Another question 
is whether
the pattern of dynamically-generated fermion mass involves the parity
violating mass ($-m\bar{\psi}_{a}\gamma^{5}\gamma^{3}\psi_{a}$) or not.
Though at this stage we do not yet know whether the dynamical symmetry 
breaking really occurs or not, in this gauge-invariant formulation of the
Thirring model, such symmetry breaking pattern is proven to be energetically
unfavorable by using the exact argument in Ref.\cite{VW}. Namely, since
the tree-level gauge action corresponding to Eq.(\ref{hls}) is real and 
positive semi-definite in Euclidean space, energetically favorable is
a parity conserving configuration consisting of half the 2-component fermions 
acquiring equal positive masses and the other half equal negative masses. 

According to the above arguments, the pattern of symmetry breaking we shall 
consider is not the parity but the chiral symmetry, i.e., of which the
breaking is $U(2N)\rightarrow U(N)\times U(N)$. Thus we investigate
the dynamical mass of the type $m\bar\psi \psi$ in the Schwinger-Dyson
equation, giving
\begin{equation}\label{sdeq}
(A(-p^2)-1)\fsl{p}-B(-p^2)=-\frac{1}{N}\int \frac{d^D q}{i(2 \pi)^D}
\gamma_{\mu}\dfrac{A(-q^2)\fsl{q}+B (-q^2)}{A^2(-q^2)q^{2}-B^{2}(-q^2)}
\Gamma_{\nu}(p,q)\;iD^{\mu \nu}(p-q),
\end{equation} 
where the full fermion propagator is written as $S(p)=i[A(-p^2)
\fsl{p}-B(-p^2)]^{-1}$, and $\Gamma_{\nu}(p,q)$ and $D_{\mu \nu}(p-q)$ 
denote the full vertex function and the full gauge boson propagator, 
respectively. Task is, by employing some appropriate approximations, to
reduce Eq.(\ref{sdeq}) to the tractable integral equation for the mass
function $M(-p^2) = B(-p^2)/A(-p^2)$. First, we here adopt the $1/N$
expansion for $\Gamma_{\nu}(p,q)$ and $D_{\mu \nu}(p-q)$ under a nonlocal
$R_{\xi}$ gauge, i.e., they are the bare vertex and the one-loop vacuum 
polarization of massless fermion loop at the $1/N$ leading order.
Then the Schwinger-Dyson equation (\ref{sdeq}) becomes the coupled integral
equations for
$A(-p^2)$ and $B(-p^2)$. They support a trivial solution $A(-p^2)=1$ and 
$B(-p^2)=0$ at the $1/N$ leading order, however, as was realized in
QED${}_{3}$, we expect to find a nonperturbative nontrivial solution
by examining them for finite $N$. 

A way is, by use of the freedom of gauge choice, to require $A(-p^{2})=1$
in a Schwinger-Dyson equation for $A(-p^{2})$. Then this gauge fulfills 
the consistency
between the bare vertex approximation and the Ward-Takahashi identity
for the hidden local $U(1)$ symmetry (or the current conservation), i.e.,
$A(0)=1$. The specific form of the gauge is determined by a Schwinger-Dyson 
integral equation for $A(-p^{2})$, and it reduces the coupled Schwinger-Dyson 
equations into a single equation for $B(-p^{2})$ which turns out to be a 
mass function, i.e., 
$M(-p^{2})=B(-p^{2})$:
\begin{equation}\label{sdeqb}
B(p^2)=\dfrac{1}{N}\int_0^{\Lambda^{D-2}}d(q^{D-2})K(p,q;G)
\dfrac{q^2 B(q^2)}{q^2+B^2(q^2)},
\end{equation}
where $\Lambda$ is ultraviolet cutoff and the kernel $K(p,q;G)$ is given by
\begin{equation}\label{ker}
K(p,q;G)={\textstyle \frac{1}{ (D-2) 2^{D-1} \pi^{(D+1)/2} \Gamma \left( 
\frac{D-1}{2}\right)}}\int_0^\pi d\theta\sin^{D-2}\theta\;d(k^2)[D-\eta(k^2)],
\end{equation}
with $k^2 = p^2 + q^2 - 2 pq \cos \theta$. 
The gauge fixing parameter $\xi$ is a function of $k^{2}$ such as
\begin{eqnarray}
\eta(k^{2})&=&\frac{-\xi(k^{2})C_{D}^{\;-1}k^{D-2}+k^{2}}{
\xi(k^{2})G^{-1}+k^{2}}\\
&=&(D-2)\left[\left(1+ \frac{Gk^{D-2}}{C_D} \right){}_{2}F_{1}
\left( \textstyle{ 1,1+\frac{D}{D-2},2+\frac{D}{D-2};-\frac{Gk^{D-2}}{C_D} }
\right)-1\right],
\end{eqnarray}
where $C_D^{-1}\equiv\frac{2\;tr\;I}{(4\pi)^{D/2}}\Gamma\left(2-
\textstyle{\frac{D}{2}}\right)B\left(\textstyle{\frac{D}{2}},\;
\textstyle{\frac{D}{2}}\right)$ and ${}_2 F_1(a,b,c;z)$ the hypergeometric 
function.
Note that the kernel $K(p,q;G)$
is positive-definite for positive arguments $p$, $q$, $G$ and is symmetric
under the exchange of $p$ and $q$.

Now let us show the existence of the nontrivial solution for the 
Schwinger-Dyson equation (\ref{sdeqb}) when $2<D<4$. Since we have in mind
the continuous phase transition, the solution of our interest is
the nontrivial solution which starts to exist without gap in the vicinity
of the phase transition point. Such a bifurcation point is identified by
the existence of an infinitesimal solution $\delta B(p^{2})$ around the trivial
solution $B(p^{2})=0$. In terms of dimensionless variables 
($p=\Lambda x^{1/(D-2)},\;\delta B(p^{2})=\Lambda\Sigma(x),\;g=G/
\Lambda^{2-D}$), Eq.(\ref{sdeqb}) is reduced to a linearized integral equation  
\begin{equation}\label{lineq}
\Sigma(x)=\frac{1}{N}\int_{\sigma_m}^{1}dyK(x^{1/(D-2)},y^{1/(D-2)};g) 
\Sigma(y),
\end{equation}
where $\sigma_{m}=(m/\Lambda)^{D-2}\;(0<\sigma_{m}\leq 1)$ is the rescaled 
infrared cutoff and in fact $m$ is nothing but the dynamically generated mass
by the normalization $m=\delta B(m^{2})$. We can rigorously prove that, 
{\it if $N$ is equal to the maximal eigenvalue of the kernel} (\ref{ker}), 
{\it then there exists a nontrivial
solution $\Sigma(x)$}. Hence, for a given $\sigma_{m}$, each line 
$N(g,\sigma_{m})$ on $(N,g)$ plane depicts a line of equal 
dynamically-generated mass $m=\Lambda\sigma^{1/(D-2)}_{m}$. Therefore, the 
critical line is 
defined by $N_{\rm cr}(g)=\lim_{\sigma_{m}\rightarrow 0}N(g,\sigma_{m})$
which separates the broken phase from the symmetric phase.
It is difficult to obtain the explicit form of the critical line $N_{\rm cr}(g)$
for arbitrary $g$, however we can get it in the limit of infinite four-fermion
coupling constant, $g\rightarrow\infty$. In this limit, the bifurcation
equation (\ref{lineq}) in (2+1)D is rewritten into a differential equation
\begin{equation}\label{difeq}
\frac{d}{dx}\Bigl(x^2 \frac{d\Sigma(x)}{dx}\Bigr)=-\frac{32}{3\pi^2 N} 
\Sigma (x),
\end{equation}
plus the infrared boundary condition $\Sigma^{\prime} (\sigma_m)=0$ and
the ultraviolet one $\mathop{[}x\Sigma^{\prime}(x)+\Sigma(x)\mathop{]}
{}_{x=1}=0$.
When $N>N_{\rm cr}\equiv 128/3\pi^2$, there is no nontrivial solution of 
Eq.(\ref{difeq}) satisfying the boundary conditions, while for $N<N_{\rm cr}$ 
the following bifurcation solutions exist:
\begin{equation}\label{eqn:sol}
\Sigma(x)=\frac{\sigma_m}{\sin(\frac{\omega}{2}\delta)}\Bigl(
\frac{x}{\sigma_m}\Bigr)^{-\frac{1}{2}}\sin\biggl\{\frac{\omega}{2} 
\Bigl[\ln\frac{x}{\sigma_m}+\delta\Bigr]\biggl\}, 
\end{equation}
where $\omega\equiv\sqrt{N_{\rm cr}/N-1}$, $\delta\equiv 2\omega^{-1}\arctan 
\omega$ and $\sigma_m$ is given by the ultraviolet boundary condition:
\begin{equation}\label{uvb}
\frac{\omega}{2}\Bigl(\ln\frac{1}{\sigma_m}+2\delta\Bigr)=n\pi,\;\;\;
\mathop{n = 1,2,}\cdots\;\;.
\end{equation}
The solution with $n=1$ is the nodeless (ground state) solution
whose scaling behavior is read from Eq.(\ref{uvb}):
\begin{equation}\label{scal}
\frac{m}{\Lambda}=e^{2 \delta}\exp\biggl[-\frac{2\pi}{\sqrt{N_{\rm cr}/N - 1}}
\biggr].
\end{equation}
The critical four-fermion number $N_{\rm cr}=128/3\pi^{2}$ is
the same as the one in QED${}_{3}$ with the nonlocal gauge.

It is turn to comment briefly on the dynamically generated mass of the gauge 
boson and the dual transformation. The vector (gauge) boson is merely an 
auxiliary field at the tree level, however it turns out to be propagating by 
obtaining the kinetic term through fermion loop effect
when the fermion acquires the dynamical mass.  
In (2+1) dimensions the pole mass $M_{V}$ of the dynamical gauge boson is given
by
\begin{equation}\label{vmass}
\frac{1}{2\pi}\biggl[\frac{4m^{2}+M_{V}^{2}}{2M_{V}}\tan^{-1}
\frac{M_{V}}{2m}-m\biggr]=G^{-1},
\end{equation}
which always satisfies a condition that $M_{V}<2m$. Once the Thirring model
is understood as a gauge theory (\ref{hls}), its effective theory can be 
expressed in terms of
dual antisymmetric-tensor field $H_{\mu_{1}\cdots\mu_{D-2}}$ of rank $D-2$,
which is actually a composite of fermions $\epsilon^{\mu_{1}\cdots\mu_{D}}
\partial_{\mu_{2}}H_{\mu_{3}\cdots\mu_{D}}=-\frac{\sqrt{G}}{(D-1)\sqrt{N}}
\sum_{a}\bar{\psi}_{a}\gamma^{\mu_{1}}\psi_{a}$. This relation 
implies the current conservation at the quantum level, which is a crucial 
property for avoiding peculiarity in the lattice study of the Thirring 
model \cite{DH}. 
In (1+1) dimensions the above relation is nothing but the one of the
bosonization, however extension to higher-dimensional case by 
integrating out the fermions leads to nonlocal bosonic Lagrangian \cite{FS}. 
In (2+1) dimensions the dual gauge field $H_{\mu}$ shares exactly the same 
pole structure with the gauge field $A_{\mu}$ irrespectively of the phase.

Analytic studies in the scheme of the Schwinger-Dyson equations with an
$1/N$ expansion yield the existence of the phase transition line on the
plane of couplings $(N,g)$ and predict the critical number of the four 
component fermion being $N_{\rm cr}=128/3\pi^{2}$ in the $g\rightarrow\infty$
limit. In order to check these intriguing questions, an appropriate systematic
method is lattice simulation. Here we present the status of the simulation
of lattice gauge theory \cite{KiKi}. The continuum Lagrangian is a $U(1)$ gauge
theory, and then the discretized version of Eq.(\ref{hls}) leads to a lattice 
gauge theory expressed in terms of a link variable $e^{i\theta_{\mu}(x)}$
and staggered fermions:
\begin{equation}\label{latac}
{\cal L}_{L}=\sum_{a}{\Phi^{\dagger}}_{a}(M^{\dagger}M)^{-1}\Phi_{a}
-N\beta\sum_{\mu}\cos{(\phi(x+\mu)-\phi(x)+\theta_{\mu}(x))},
\end{equation}
where $\beta=1/g$ and 
\begin{equation}
M_{x,y}=\frac{1}{2}\sum_{\mu}\eta_{\mu}(x)(e^{i\theta_{\mu}(x)}\delta_{y,x+\mu}
-e^{-i\theta_{\mu}(x-\mu)}\delta_{y,x-\mu})+m\delta_{x,y}.
\end{equation}
Note that one can not discriminate whether this lattice Lagrangian (\ref{latac})
at the tree level lies in the compact formulation or in the noncompact one since
there is no gauge kinetic term at the tree level. This issue is rather subtle 
because each continuum limit may arrives at different continuum theory as has 
been done in QED${}_{3}$.

So far, simulations for $N=2, 4$ and 6 systems on a $8^{3}$ lattice volume
with bare fermion mass $m=0.05, 0.025$ and 0.0125 are completed by use of 
hybrid Monte Carlo algorithm and preliminary data on a $16^3$ lattice for
$N=2$ and 6 are prepared for finite volume effect study \cite{KiKi}. 
Numerical simulation results on a $8^3$ lattice are 
summarized in Fig. 1 where the plot points
are those obtained by $m\rightarrow0$ extrapolation.
$N = 2$ shows smooth change of the chiral condensate and a
long tail into weak coupling regime, $N = 4$ shows a complex
behavior of the condensate near the critical point, and $N = 6$
shows an abrupt change of the condensate which can be interpreted as a mean 
field type phase transition.
The result from a coarse-grained lattice alludes to the existence of the
critical fermion number $N_{{\rm cr}}$ $(2<N_{{\rm cr}}<6)$ for which the
chiral behavior changes its character. 
It is consistent with the result in the continuum Thirring model reformulated
as a gauge theory. Preliminary as the result is, the above suggests
possible existence of the critical $N_{\rm cr}$ in the lattice gauge
formulation of the Thirring model. 
Data on a $16^3$ lattice tells us that, though the dependence of the chiral
condensate on both the size of the lattices and the extrapolation method is
considerable quantitatively (particularly in the very weak coupling region
of $N=2$ case), the overall qualitative nature of the chiral condensate does
not seem to be changed.
It is interesting to wait further lattice result of $N=6$ case
based on the auxiliary vecotr field \cite{DH} for the comparison with
the above \cite{KiKi}.

\vspace{7mm}


\setlength{\unitlength}{0.240900pt}
\ifx\plotpoint\undefined\newsavebox{\plotpoint}\fi
\sbox{\plotpoint}{\rule[-0.200pt]{0.400pt}{0.400pt}}%
\begin{picture}(1500,900)(0,0)
\font\gnuplot=cmr10 at 10pt
\gnuplot
\sbox{\plotpoint}{\rule[-0.200pt]{0.400pt}{0.400pt}}%
\put(70,468){\makebox(0,0)[r]{$\bigl<\bar{\psi}\psi\bigr>$}}
\put(820,-50){\makebox(0,0)[r]{$\beta$}}
\put(176.0,169.0){\rule[-0.200pt]{303.534pt}{0.400pt}}
\put(176.0,68.0){\rule[-0.200pt]{0.400pt}{194.888pt}}
\put(176.0,68.0){\rule[-0.200pt]{4.818pt}{0.400pt}}
\put(154,68){\makebox(0,0)[r]{-0.1}}
\put(1416.0,68.0){\rule[-0.200pt]{4.818pt}{0.400pt}}
\put(176.0,169.0){\rule[-0.200pt]{4.818pt}{0.400pt}}
\put(154,169){\makebox(0,0)[r]{0}}
\put(1416.0,169.0){\rule[-0.200pt]{4.818pt}{0.400pt}}
\put(176.0,270.0){\rule[-0.200pt]{4.818pt}{0.400pt}}
\put(154,270){\makebox(0,0)[r]{0.1}}
\put(1416.0,270.0){\rule[-0.200pt]{4.818pt}{0.400pt}}
\put(176.0,371.0){\rule[-0.200pt]{4.818pt}{0.400pt}}
\put(154,371){\makebox(0,0)[r]{0.2}}
\put(1416.0,371.0){\rule[-0.200pt]{4.818pt}{0.400pt}}
\put(176.0,473.0){\rule[-0.200pt]{4.818pt}{0.400pt}}
\put(154,473){\makebox(0,0)[r]{0.3}}
\put(1416.0,473.0){\rule[-0.200pt]{4.818pt}{0.400pt}}
\put(176.0,574.0){\rule[-0.200pt]{4.818pt}{0.400pt}}
\put(154,574){\makebox(0,0)[r]{0.4}}
\put(1416.0,574.0){\rule[-0.200pt]{4.818pt}{0.400pt}}
\put(176.0,675.0){\rule[-0.200pt]{4.818pt}{0.400pt}}
\put(154,675){\makebox(0,0)[r]{0.5}}
\put(1416.0,675.0){\rule[-0.200pt]{4.818pt}{0.400pt}}
\put(176.0,776.0){\rule[-0.200pt]{4.818pt}{0.400pt}}
\put(154,776){\makebox(0,0)[r]{0.6}}
\put(1416.0,776.0){\rule[-0.200pt]{4.818pt}{0.400pt}}
\put(176.0,877.0){\rule[-0.200pt]{4.818pt}{0.400pt}}
\put(154,877){\makebox(0,0)[r]{0.7}}
\put(1416.0,877.0){\rule[-0.200pt]{4.818pt}{0.400pt}}
\put(176.0,68.0){\rule[-0.200pt]{0.400pt}{4.818pt}}
\put(176,23){\makebox(0,0){0}}
\put(176.0,857.0){\rule[-0.200pt]{0.400pt}{4.818pt}}
\put(309.0,68.0){\rule[-0.200pt]{0.400pt}{4.818pt}}
\put(309,23){\makebox(0,0){0.2}}
\put(309.0,857.0){\rule[-0.200pt]{0.400pt}{4.818pt}}
\put(441.0,68.0){\rule[-0.200pt]{0.400pt}{4.818pt}}
\put(441,23){\makebox(0,0){0.4}}
\put(441.0,857.0){\rule[-0.200pt]{0.400pt}{4.818pt}}
\put(574.0,68.0){\rule[-0.200pt]{0.400pt}{4.818pt}}
\put(574,23){\makebox(0,0){0.6}}
\put(574.0,857.0){\rule[-0.200pt]{0.400pt}{4.818pt}}
\put(707.0,68.0){\rule[-0.200pt]{0.400pt}{4.818pt}}
\put(707,23){\makebox(0,0){0.8}}
\put(707.0,857.0){\rule[-0.200pt]{0.400pt}{4.818pt}}
\put(839.0,68.0){\rule[-0.200pt]{0.400pt}{4.818pt}}
\put(839,23){\makebox(0,0){1}}
\put(839.0,857.0){\rule[-0.200pt]{0.400pt}{4.818pt}}
\put(972.0,68.0){\rule[-0.200pt]{0.400pt}{4.818pt}}
\put(972,23){\makebox(0,0){1.2}}
\put(972.0,857.0){\rule[-0.200pt]{0.400pt}{4.818pt}}
\put(1104.0,68.0){\rule[-0.200pt]{0.400pt}{4.818pt}}
\put(1104,23){\makebox(0,0){1.4}}
\put(1104.0,857.0){\rule[-0.200pt]{0.400pt}{4.818pt}}
\put(1237.0,68.0){\rule[-0.200pt]{0.400pt}{4.818pt}}
\put(1237,23){\makebox(0,0){1.6}}
\put(1237.0,857.0){\rule[-0.200pt]{0.400pt}{4.818pt}}
\put(1370.0,68.0){\rule[-0.200pt]{0.400pt}{4.818pt}}
\put(1370,23){\makebox(0,0){1.8}}
\put(1370.0,857.0){\rule[-0.200pt]{0.400pt}{4.818pt}}
\put(176.0,68.0){\rule[-0.200pt]{303.534pt}{0.400pt}}
\put(1436.0,68.0){\rule[-0.200pt]{0.400pt}{194.888pt}}
\put(176.0,877.0){\rule[-0.200pt]{303.534pt}{0.400pt}}
\put(176.0,68.0){\rule[-0.200pt]{0.400pt}{194.888pt}}
\put(342,793){\raisebox{-.8pt}{\makebox(0,0){$\Diamond$}}}
\put(508,764){\raisebox{-.8pt}{\makebox(0,0){$\Diamond$}}}
\put(673,634){\raisebox{-.8pt}{\makebox(0,0){$\Diamond$}}}
\put(707,568){\raisebox{-.8pt}{\makebox(0,0){$\Diamond$}}}
\put(773,448){\raisebox{-.8pt}{\makebox(0,0){$\Diamond$}}}
\put(839,313){\raisebox{-.8pt}{\makebox(0,0){$\Diamond$}}}
\put(905,229){\raisebox{-.8pt}{\makebox(0,0){$\Diamond$}}}
\put(1005,191){\raisebox{-.8pt}{\makebox(0,0){$\Diamond$}}}
\put(1038,183){\raisebox{-.8pt}{\makebox(0,0){$\Diamond$}}}
\put(1104,176){\raisebox{-.8pt}{\makebox(0,0){$\Diamond$}}}
\put(1171,173){\raisebox{-.8pt}{\makebox(0,0){$\Diamond$}}}
\put(1337,173){\raisebox{-.8pt}{\makebox(0,0){$\Diamond$}}}
\put(342.0,790.0){\rule[-0.200pt]{0.400pt}{1.445pt}}
\put(332.0,790.0){\rule[-0.200pt]{4.818pt}{0.400pt}}
\put(332.0,796.0){\rule[-0.200pt]{4.818pt}{0.400pt}}
\put(508.0,762.0){\rule[-0.200pt]{0.400pt}{1.204pt}}
\put(498.0,762.0){\rule[-0.200pt]{4.818pt}{0.400pt}}
\put(498.0,767.0){\rule[-0.200pt]{4.818pt}{0.400pt}}
\put(673.0,632.0){\rule[-0.200pt]{0.400pt}{1.204pt}}
\put(663.0,632.0){\rule[-0.200pt]{4.818pt}{0.400pt}}
\put(663.0,637.0){\rule[-0.200pt]{4.818pt}{0.400pt}}
\put(707.0,566.0){\rule[-0.200pt]{0.400pt}{1.204pt}}
\put(697.0,566.0){\rule[-0.200pt]{4.818pt}{0.400pt}}
\put(697.0,571.0){\rule[-0.200pt]{4.818pt}{0.400pt}}
\put(773.0,446.0){\rule[-0.200pt]{0.400pt}{0.964pt}}
\put(763.0,446.0){\rule[-0.200pt]{4.818pt}{0.400pt}}
\put(763.0,450.0){\rule[-0.200pt]{4.818pt}{0.400pt}}
\put(839.0,312.0){\rule[-0.200pt]{0.400pt}{0.723pt}}
\put(829.0,312.0){\rule[-0.200pt]{4.818pt}{0.400pt}}
\put(829.0,315.0){\rule[-0.200pt]{4.818pt}{0.400pt}}
\put(905.0,228.0){\rule[-0.200pt]{0.400pt}{0.482pt}}
\put(895.0,228.0){\rule[-0.200pt]{4.818pt}{0.400pt}}
\put(895.0,230.0){\rule[-0.200pt]{4.818pt}{0.400pt}}
\put(1005.0,190.0){\rule[-0.200pt]{0.400pt}{0.482pt}}
\put(995.0,190.0){\rule[-0.200pt]{4.818pt}{0.400pt}}
\put(995.0,192.0){\rule[-0.200pt]{4.818pt}{0.400pt}}
\put(1038.0,182.0){\rule[-0.200pt]{0.400pt}{0.482pt}}
\put(1028.0,182.0){\rule[-0.200pt]{4.818pt}{0.400pt}}
\put(1028.0,184.0){\rule[-0.200pt]{4.818pt}{0.400pt}}
\put(1104.0,176.0){\usebox{\plotpoint}}
\put(1094.0,176.0){\rule[-0.200pt]{4.818pt}{0.400pt}}
\put(1094.0,177.0){\rule[-0.200pt]{4.818pt}{0.400pt}}
\put(1171.0,172.0){\rule[-0.200pt]{0.400pt}{0.482pt}}
\put(1161.0,172.0){\rule[-0.200pt]{4.818pt}{0.400pt}}
\put(1161.0,174.0){\rule[-0.200pt]{4.818pt}{0.400pt}}
\put(1337.0,172.0){\rule[-0.200pt]{0.400pt}{0.482pt}}
\put(1327.0,172.0){\rule[-0.200pt]{4.818pt}{0.400pt}}
\put(1327.0,174.0){\rule[-0.200pt]{4.818pt}{0.400pt}}
\put(229,664){\makebox(0,0){$+$}}
\put(242,671){\makebox(0,0){$+$}}
\put(309,641){\makebox(0,0){$+$}}
\put(342,606){\makebox(0,0){$+$}}
\put(375,526){\makebox(0,0){$+$}}
\put(408,370){\makebox(0,0){$+$}}
\put(441,221){\makebox(0,0){$+$}}
\put(474,184){\makebox(0,0){$+$}}
\put(508,177){\makebox(0,0){$+$}}
\put(541,174){\makebox(0,0){$+$}}
\put(673,171){\makebox(0,0){$+$}}
\put(839,170){\makebox(0,0){$+$}}
\put(1005,170){\makebox(0,0){$+$}}
\put(1171,169){\makebox(0,0){$+$}}

\put(219.00,661.00){\rule[-0.200pt]{4.818pt}{0.400pt}}
\put(219.00,666.00){\rule[-0.200pt]{4.818pt}{0.400pt}}

\put(232.00,668.00){\rule[-0.200pt]{4.818pt}{0.400pt}}
\put(232.00,673.00){\rule[-0.200pt]{4.818pt}{0.400pt}}

\put(299.00,638.00){\rule[-0.200pt]{4.818pt}{0.400pt}}
\put(299.00,643.00){\rule[-0.200pt]{4.818pt}{0.400pt}}

\put(332.00,604.00){\rule[-0.200pt]{4.818pt}{0.400pt}}
\put(332.00,609.00){\rule[-0.200pt]{4.818pt}{0.400pt}}

\put(365.00,524.00){\rule[-0.200pt]{4.818pt}{0.400pt}}
\put(365.00,529.00){\rule[-0.200pt]{4.818pt}{0.400pt}}

\put(398.00,368.00){\rule[-0.200pt]{4.818pt}{0.400pt}}
\put(398.00,372.00){\rule[-0.200pt]{4.818pt}{0.400pt}}

\put(431.00,219.00){\rule[-0.200pt]{4.818pt}{0.400pt}}
\put(431.00,222.00){\rule[-0.200pt]{4.818pt}{0.400pt}}

\put(464.00,183.00){\rule[-0.200pt]{4.818pt}{0.400pt}}
\put(464.00,185.00){\rule[-0.200pt]{4.818pt}{0.400pt}}

\put(498.00,177.00){\rule[-0.200pt]{4.818pt}{0.400pt}}
\put(498.00,178.00){\rule[-0.200pt]{4.818pt}{0.400pt}}

\put(531.00,173.00){\rule[-0.200pt]{4.818pt}{0.400pt}}
\put(531.00,174.00){\rule[-0.200pt]{4.818pt}{0.400pt}}

\put(663.00,170.00){\rule[-0.200pt]{4.818pt}{0.400pt}}
\put(663.00,171.00){\rule[-0.200pt]{4.818pt}{0.400pt}}

\put(829.00,169.00){\rule[-0.200pt]{4.818pt}{0.400pt}}
\put(829.00,170.00){\rule[-0.200pt]{4.818pt}{0.400pt}}

\put(995.00,169.00){\rule[-0.200pt]{4.818pt}{0.400pt}}
\put(995.00,170.00){\rule[-0.200pt]{4.818pt}{0.400pt}}

\put(1161.00,168.00){\rule[-0.200pt]{4.818pt}{0.400pt}}
\put(1161.00,169.00){\rule[-0.200pt]{4.818pt}{0.400pt}}

\sbox{\plotpoint}{\rule[-0.400pt]{0.800pt}{0.800pt}}%
\put(229,507){\raisebox{-.8pt}{\makebox(0,0){$\Box$}}}
\put(242,498){\raisebox{-.8pt}{\makebox(0,0){$\Box$}}}
\put(275,448){\raisebox{-.8pt}{\makebox(0,0){$\Box$}}}
\put(282,407){\raisebox{-.8pt}{\makebox(0,0){$\Box$}}}
\put(289,365){\raisebox{-.8pt}{\makebox(0,0){$\Box$}}}
\put(295,306){\raisebox{-.8pt}{\makebox(0,0){$\Box$}}}
\put(302,230){\raisebox{-.8pt}{\makebox(0,0){$\Box$}}}
\put(309,187){\raisebox{-.8pt}{\makebox(0,0){$\Box$}}}
\put(322,166){\raisebox{-.8pt}{\makebox(0,0){$\Box$}}}
\put(342,171){\raisebox{-.8pt}{\makebox(0,0){$\Box$}}}
\put(375,172){\raisebox{-.8pt}{\makebox(0,0){$\Box$}}}
\put(441,170){\raisebox{-.8pt}{\makebox(0,0){$\Box$}}}
\put(508,169){\raisebox{-.8pt}{\makebox(0,0){$\Box$}}}
\put(673,170){\raisebox{-.8pt}{\makebox(0,0){$\Box$}}}
\put(839,169){\raisebox{-.8pt}{\makebox(0,0){$\Box$}}}
\put(229.0,504.0){\rule[-0.400pt]{0.800pt}{1.204pt}}
\put(219.0,504.0){\rule[-0.400pt]{4.818pt}{0.800pt}}
\put(219.0,509.0){\rule[-0.400pt]{4.818pt}{0.800pt}}
\put(242.0,495.0){\rule[-0.400pt]{0.800pt}{1.204pt}}
\put(232.0,495.0){\rule[-0.400pt]{4.818pt}{0.800pt}}
\put(232.0,500.0){\rule[-0.400pt]{4.818pt}{0.800pt}}
\put(275.0,446.0){\rule[-0.400pt]{0.800pt}{0.964pt}}
\put(265.0,446.0){\rule[-0.400pt]{4.818pt}{0.800pt}}
\put(265.0,450.0){\rule[-0.400pt]{4.818pt}{0.800pt}}
\put(282.0,405.0){\rule[-0.400pt]{0.800pt}{0.964pt}}
\put(272.0,405.0){\rule[-0.400pt]{4.818pt}{0.800pt}}
\put(272.0,409.0){\rule[-0.400pt]{4.818pt}{0.800pt}}
\put(289.0,363.0){\rule[-0.400pt]{0.800pt}{0.964pt}}
\put(279.0,363.0){\rule[-0.400pt]{4.818pt}{0.800pt}}
\put(279.0,367.0){\rule[-0.400pt]{4.818pt}{0.800pt}}
\put(295.0,304.0){\rule[-0.400pt]{0.800pt}{0.964pt}}
\put(285.0,304.0){\rule[-0.400pt]{4.818pt}{0.800pt}}
\put(285.0,308.0){\rule[-0.400pt]{4.818pt}{0.800pt}}
\put(302.0,229.0){\usebox{\plotpoint}}
\put(292.0,229.0){\rule[-0.400pt]{4.818pt}{0.800pt}}
\put(292.0,232.0){\rule[-0.400pt]{4.818pt}{0.800pt}}
\put(309.0,186.0){\usebox{\plotpoint}}
\put(299.0,186.0){\rule[-0.400pt]{4.818pt}{0.800pt}}
\put(299.0,188.0){\rule[-0.400pt]{4.818pt}{0.800pt}}
\put(322.0,165.0){\usebox{\plotpoint}}
\put(312.0,165.0){\rule[-0.400pt]{4.818pt}{0.800pt}}
\put(312.0,167.0){\rule[-0.400pt]{4.818pt}{0.800pt}}
\put(342.0,170.0){\usebox{\plotpoint}}
\put(332.0,170.0){\rule[-0.400pt]{4.818pt}{0.800pt}}
\put(332.0,171.0){\rule[-0.400pt]{4.818pt}{0.800pt}}
\put(375.0,171.0){\usebox{\plotpoint}}
\put(365.0,171.0){\rule[-0.400pt]{4.818pt}{0.800pt}}
\put(365.0,173.0){\rule[-0.400pt]{4.818pt}{0.800pt}}
\put(441.0,169.0){\usebox{\plotpoint}}
\put(431.0,169.0){\rule[-0.400pt]{4.818pt}{0.800pt}}
\put(431.0,170.0){\rule[-0.400pt]{4.818pt}{0.800pt}}
\put(508.0,169.0){\usebox{\plotpoint}}
\put(498.0,169.0){\rule[-0.400pt]{4.818pt}{0.800pt}}
\put(498.0,170.0){\rule[-0.400pt]{4.818pt}{0.800pt}}
\put(673.0,170.0){\usebox{\plotpoint}}
\put(663.0,170.0){\rule[-0.400pt]{4.818pt}{0.800pt}}
\put(663.0,171.0){\rule[-0.400pt]{4.818pt}{0.800pt}}
\put(839.0,169.0){\usebox{\plotpoint}}
\put(829.0,169.0){\rule[-0.400pt]{4.818pt}{0.800pt}}
\put(829.0,170.0){\rule[-0.400pt]{4.818pt}{0.800pt}}
\end{picture}

\vspace{6mm}

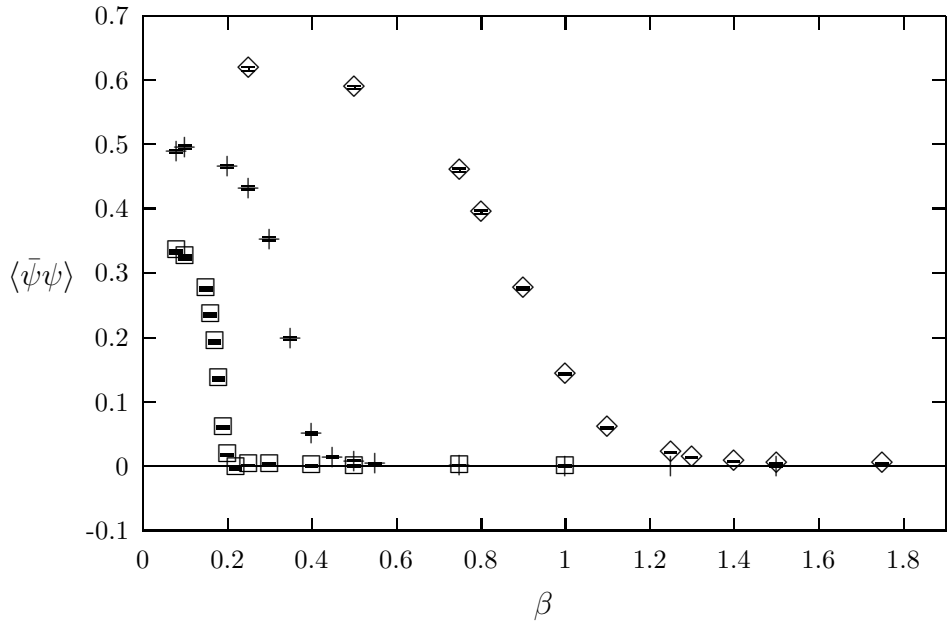
Figure 1: The chiral condensate, $\bigl<\bar{\psi}\psi\bigr>$, vs. $\beta$
for $N=2$ ($\Diamond$ points), $N=4$ ($+$ points), and $N=6$ ($\Box$ points).

\vspace{6mm}

In this report, we have reviewed how to reformulate the Thirring model as
a gauge theory both on continuum and lattice, and, in such context, the
dynamical symmetry breaking for $N$ four component Dirac fermions in (2+1)D.
The Schwinger-Dyson approach in cooperation with $1/N$ expansion reaches
a rigorous proof of the existence of the chiral phase transition at a certain 
number of $N$ and $g$ under the light shed by the hidden local symmetry.
The application of Vafa-Witten theorem forces the parity preservation
in the dynamically-generated fermion mass spectra and the argument of 
parity-violating anomaly prevents the parity violation in gauge sector.
In the infinite coupling limit $(g\rightarrow\infty)$, the Schwinger-Dyson
equation was solved analytically and yields the critical fermion number
$N_{\rm cr}=128/3\pi^{2}$ in perfect agreement with $N_{\rm cr}$ in QED${}_{3}$.
More systematic analysis of the Schwinger-Dyson equations is in progess 
\cite{SSY}.
According to the present lattice simulations on $8^3$ and $16^3$ lattices, 
the lattice
gauge formulation seems to support the existence of $N_{\rm cr}$ 
$(2<N_{\rm cr}<6)$, which should be tested again by further lattice 
simulation \cite{KKS}.

\vspace{5mm}

{\Large{\bf\noindent Acknowledgments}}

\vspace{5mm}
We would like to thank Seyong Kim for helpful discussions on the lattice
gauge theories.
Y.K.'s research was also supported in part by the KOSEF(Brain Pool Program, 
95-0702-04-01-3, CTP in Seoul National University) and the Korean Ministry 
of Education (BSRI-95-2413). 
K.Y.'s research is supported by the Sumitomo Foundation and a Grant-in Aid
for Scientific Research from the Ministry of Education, Science and Culture
(No. 05640339).

\end{document}